\documentclass[aps,prl,superscriptaddress,showpacs]{revtex4}

\usepackage{amsmath}
\usepackage{amsfonts}
\usepackage{bm}

\ifx\pdfoutput\undefined
        \usepackage{graphicx}
\else
        \usepackage[pdftex]{graphicx}
\fi

\begin{document}

\title{Pulse Propagation in Resonant  Tunneling} 

\author{Ulrich Wulf}
\affiliation{Technische Universit\"at Cottbus, 
Lehrstuhl f\"ur Theoretische Physik and IHP/BTU Joint Lab, Postfach
101344,
              03013 Cottbus, Germany }

\author{V. V. Skalozub}
\affiliation{Dniepropetrovsk National University, Dniepropetrovsk 49050,
Ukraine}

\date{\today}

\begin{abstract}
We consider the analytically solvable model of a Gaussian pulse
tunneling through a transmission resonance with a Breit-Wigner
characteristic.
The solution
allows for the identification  of two opposite  pulse propagation regimes:
if the resonance is broad compared to  the energetic width of
the incident Gaussian pulse
 a weakly deformed and slightly delayed transmitted Gaussian pulse is found.
In the opposite limit of a narrow resonance the dying out
of the transmitted pulse is dominated by the slow exponential
decay characteristic of a quasi-bound state with a long life time
(decaying state). We discuss the limitation
of the achievable pulse transfer rate resulting from
the slow decay. 
Finally, it is demonstrated that
for narrow resonances
a small second component is superimposed to the exponential
decay which leads to characteristic
interference oscillations.
\end{abstract}


\maketitle 

\section{introduction}
The transmission of pulses in tunneling transport
is a basic problem in quantum mechanics\cite{hauge,mizuta,monsen}.
Here the case of an incident
wave packet transmitted through a resonant
quasi-bound level of a quantum system (QS) is 
a fundamental issue under
current study
\cite{villa,konsek,pereyra1,calderon1,grossel,pereyra2,stovneng}.
In our previous paper\cite{roxana} an S-matrix representation
of the quasi-bound resonance levels of a given QS
has been derived: depending on the complex asymmetry
parameter a Breit-Wigner or a general Fano resonance line results.
In our study we obtain analytical
solutions for the case of an incident Gaussian wave packet
interacting with a Breit-Wigner resonance.
Here we focus on the simplest
case when the central position of the incident
wave packet in the momentum space coincides with that of the resonance level.
Our analytical model allows for a continuous transition between
the limits of a broad and a narrow resonance
on scale of the energetic width of the incident packet:
in the presence of a
broad resonance only a weak deformation and a small delay
of the incident Gaussian pulse is found\cite{levin}.
In the limit of a narrow resonance the signal is dominated
by a decaying state in the $QS$\cite{stovneng}. 
We study the shape of the transmitted pulse in this regime:
the dropping flank of the transmitted
pulse is not Gaussian any more but dominated
by a slow exponential time characteristic
of the decaying state.
In addition, there is a weak second
component in the transmitted signal
representing a small deformed Gaussian pulse.
Its superposition with the primary exponentially decaying component
causes characteristic interference oscillations for which we 
give an explicit expression.
\begin{figure}[b]
\begin{center}
\includegraphics*[width=3.in]{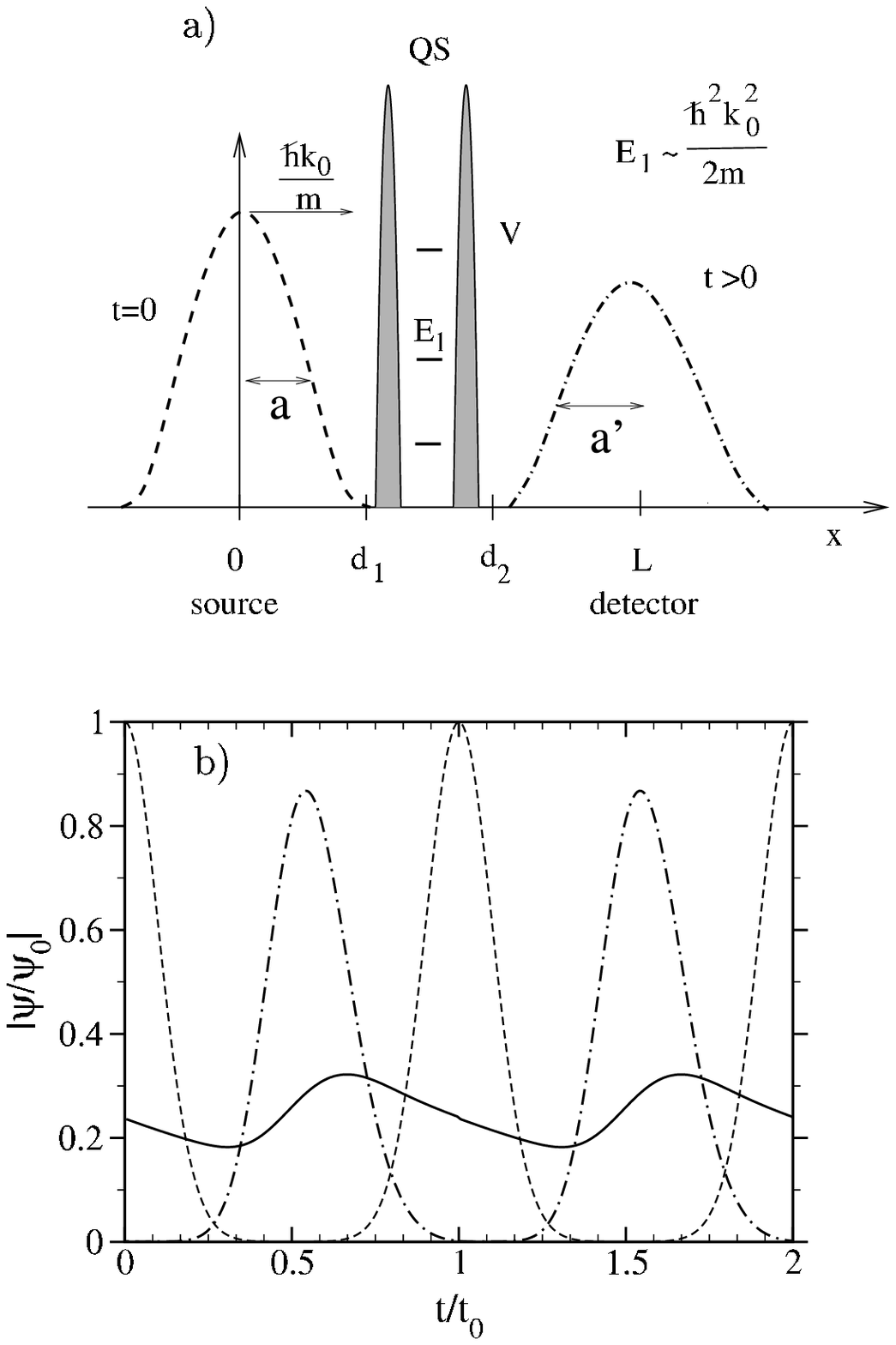}
\end{center}
\caption{ Part (a):
Schematic representation of the incident Gaussian packet (dashed line)
approaching the QS.
The transmitted signal (dash-dotted line) is detected at 
$x=L$. \\
Part (b): Absolute value of the wave function versus time. 
In dashed line the sequence of the incident 
pulses at $x = 0$. 
In dash-dotted line
the quasi-stationary limit for the transmitted pulses detected at $x=L$
for $\rho=2$ (broad resonance) and in solid  line for $\rho = 0.1$
(narrow resonance). 
The further parameters are $L = 5 a$, 
$k_0 a= k_1 a = 10$, and $|S_0|=1$.}
\label{Fig0}
\end{figure}

\section{model}
We analyze a set-up as depicted in Fig.\ \ref{Fig0}:
an  incident standard Gaussian wave packet 
\begin{equation}
\psi_0(x,t) =
{\psi_0 \over \sqrt{1 + i{\hbar \over m a^2 } t }}
\exp{ \left[ - {(x-v_0t)^2 \over  2 a^2 (1 + i {\hbar \over m a^2} t) } \right]}
\exp{\left[i k_0 x - i {\hbar k_0^2 \over 2m} t \right]}. 
\label{psio}
\end{equation}
is approaching  a tunneling barrier with a group velocity
 $v_0 = \hbar k_0/m$. At $t=0$ the incident packet has a width of $a$
in real space.
The transmitted pulse is given by the expression
\begin{equation}
\psi(x > d_2  ,t) = \int {dk \over \sqrt{2\pi} }  \psi_0(k)
S(k) \exp{ \left[i kx -  i {\hbar \over 2m} k^2 t \right]},
\label{genera}
\end{equation}
where $\psi_0(k)$ is the Fourier transform of $\psi_0(x,t=0)$.
A short range scattering potential is assumed 
so that $V(x)=0$ outside the interval $d_1 \leq x \leq d_2$.
Furthermore, it is assumed that there is
an isolated  resonance at $ E_1 \approx \hbar^2k_0^2 /(2m) $,
i. e. close to the mean kinetic energy of the incident pulse.
Then in the range of finite $|\psi_0 (k)|$ the transmission coefficient
can be approximated by a Breit-Wigner characteristic
\begin{equation}
S(k)= S_0 { i \frac{\Gamma_k}{2}  \over k - k_1 + i \frac{\Gamma_k}{2}}.
\label{sk}
\end{equation}
From Eqs.\ (\ref{psio}) - (\ref{sk}) we find with 
the help of Ref.\ \cite{gradstein}
\begin{equation}
\psi(x > 0 ,t)  = \psi_0 S_0 \rho
 \sqrt{\pi\over 2}
               \exp{ (- \gamma^2 \beta -i q \gamma) }
  \mbox{erfc} (z) 
\exp{\left[i k_0 x - i {\hbar k_0^2 \over 2m} t \right]},
\label{centra}
\end{equation}
where  $\mbox{erfc}$ is the complementary error function
taken at the argument
\begin{equation}
z = \left( {q \over 2 \sqrt{\beta }}
-i \gamma \sqrt{\beta} \right),
\label{zvalue}
\end{equation}  
with
$q = (x - v_0 t)/a$, $\beta = (1 + i \tau)/2$, $\tau = t/ t_0$,
$t_0 =  m a^2/\hbar$,
and  $\gamma = a[k_0 - k_1 + i\Gamma_k /2 ] \equiv
\Delta + i \rho $.

\section{Results}

The different pulse propagation regimes are illustrated in 
Fig. \ \ref{Fig0}(b): a sequence of 
well separated incident Gaussian pulses is created at
integer $t/t_0=n$ with the center in real space at $x=0$ (source).
For the broad enough resonance
with $\rho >1$ the signal detected at $x = L$ is a 
sequence of well separated weakly deformed Gaussian pulses.
These pulses arrive
with a small delay with respect to the times $t/t_0 = n + 1/2$ 
at which the
signal without scattering potential would arrive.
In contrast, if the  resonance is  narrow, $\rho \ll 1$,
 the
transmitted pulses are strongly weakened and  deformed so that they are
 not separable any more. As we will show below
this deterioration of the signal is caused by the exponential
time  characteristic of a decaying state with a
life time exceeding the time interval of two subsequent
pulses.
The parameters used for the calculations in Fig.\ \ref{Fig0}(b) have been 
chosen to demonstrate how a nearly optimal signal transfer can be achieved
for a broad resonance: 
i.) the detector  is positioned in a minimum distance of
$L \approx 5 \div 10 a$ from the source. 
As illustrated in Fig.\ \ref{Fig0}(a) this choice
results from the requirements, first, that when the pulse is prepared
or detected it should be well separated from
the resonant tunneling structure and, second, that the spatial width of
the signal should be comparable  to the size of
the tunneling structure. ii.) the traveling time of the signal
should not exceed the characteristic time $t_0 = ma^2 /\hbar $
of the quantum spread 
of the free wave packet in Eq.\ (\ref{psio}).
For an electron packet with a width of $a=10nm$ this time
is very short, $t_0 \approx 10^{-12}s$. 
We can therefore define a minimum 
speed $v_{min} = L / t_0 = \hbar  k_{0;min} /m$
(we neglect here the spread
induced by the Coulomb interaction). iii.)
for a maximum signal we require optimal resonance so that $k_0 = k_1$
($\Delta=0$).
Together with the second requirement
we then obtain a minimum energy of the resonance level of 
$E_1 \approx \hbar^2 k_{0;min}^2 / (2m)$ 
which is for $a=10nm$ in the range of a few $meV$. 
iv.) the imaginary part of the resonance energy is fixed by the
conditions of $\rho > 1$ with the constraint that the resonance
should still be  isolated. 
v.) for an effectively symmetrical
double barrier structure the absolute value of $S_0$ is very close to
unity\cite{paul}.

\begin{figure}[t]
\begin{center}
\noindent \includegraphics*[width=3.5in]{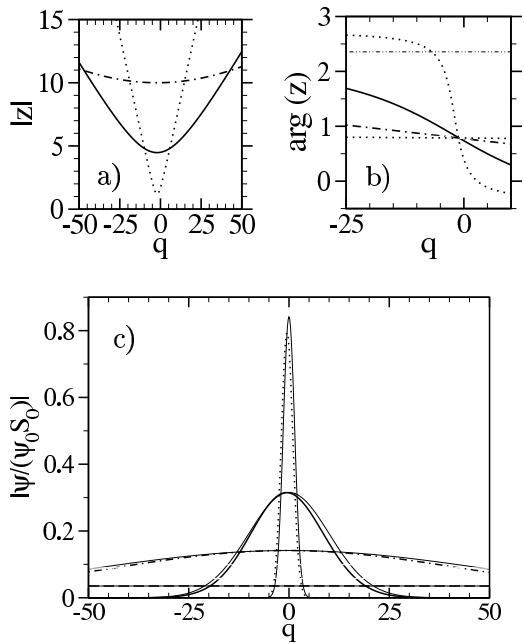}
\end{center}
\caption{Part (c): Absolute value of the transmitted wave function  for
a broad resonance
 $\rho =2.0$ at $\tau = 1.0$ (dotted line)
 $\tau = 10$ (solid line), $\tau =50$ (dash-dotted line),
and $\tau = 800$ (dotted line). Thick lines represent the exact result
in Eq.\ (\protect\ref{centra})
and, for comparison,
thin solid lines the result for $\psi = \psi_0(x,t)$
in Eq.\ (\protect\ref{psio}).
For $\tau = 800$ the corresponding
lines are not resolvable.
Part (a) absolute value of $z$
(see Eq.\ (\protect \ref{zvalue})) and
part (b) arg($z$) with the same line-coding as in part (a).
The  thin dash-dot-dotted
line in $(b)$ represents $\mbox{arg}(z) = 3\pi/4$.}
\label{Fig2}
\end{figure}

\begin{figure}[b]
\begin{center}
\noindent \includegraphics*[width=4.7in]{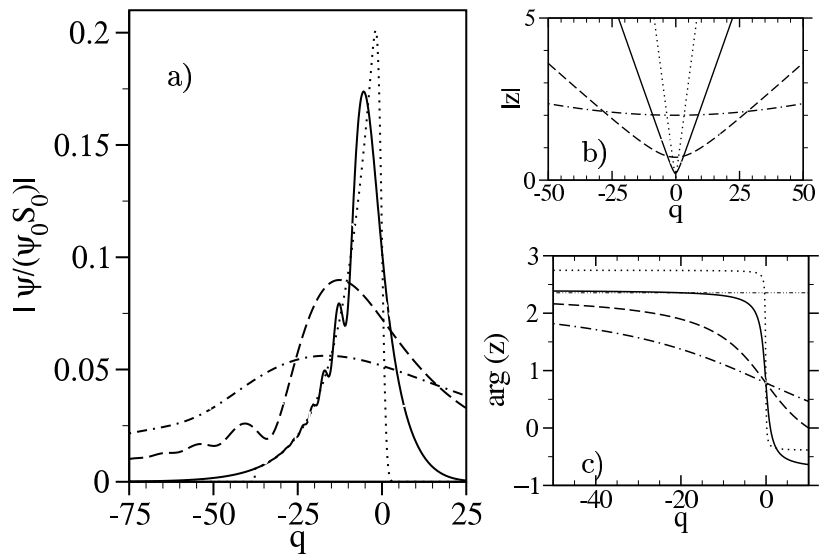}
\end{center}
\caption{(a) Absolute value of the transmitted wave function,
(b) absolute value of $z$,  and (c) the argument of $z$
at $\rho =.1$, $l_1 = l_0$ $\tau = 1.0$ (dotted line),
and
 $\tau = 10.$ (solid line), $\tau = 100$ (dashed line),
and $\tau = 300$ (dash-dotted line). The  thin dash-dot-dotted
line in $c$ represents $\mbox{arg}(z) = 3\pi/4$.}
\label{Fig3}
\end{figure}

For comparison we exemplify 
 the cases of a broad 
and a narrow resonance
in  Fig.\  \ref{Fig2} and Fig.\ \ref{Fig3}, respectively. 
The basic differences can be 
understood in a proper expansion of the complementary 
error function in Eq.\ (\ref{centra}).

For a broad resonance $\rho$ is large and
we find for the relevant values of 
$q$ (where $|\psi|$ deviates significantly from zero)
that  the argument of $z$ is less than $3\pi/4$ (see
Fig.\ \ref{Fig2} (b)). 
Under this condition
we can introduce an asymptotic $z$ expansion of the error function 
as given in \cite{abramowitz}  yielding
\begin{equation}
\psi(x,t) =
S_0 {\rho \sqrt{\beta} \over z}
\left[1 +\sum_{m=1} {a_m \over (2z^2)^m} \right]
\psi_0(x,t) \equiv \psi_{DG}(x,t),
\label{appro1}
\end{equation}
with $a_m = (-1)^m [1 \times 3 \times 5  ...\times(2m-1)]$.
The expression in Eq.\ (\ref{appro1}) 
describes a weakly distorted  Gaussian (DG) pulse.
The relatively weak distortion of the incident
Gaussian pulse $\psi_0(x,t)$ follows, first, from the fact that at
large $\rho$
the parameter $1/(2|z|^2)$ is small compared to unity. 
Therefore, the
second term in the square bracket in Eq.\ (\ref{appro1})
gives only a small correction.
In fact, only the first two terms in the sum over $m$ in  
Eq.\ (\ref{appro1}) are necessary
to approximate the analytical result from Eq.\ (\ref{centra})
within plot resolution.
Second, for large $\rho$ the argument $z$ depends only very
weakly on $q$. This  explains 
that the factor $1/z$ in front of the square
bracket in Eq.\ (\ref{appro1}) merely 
leads to  a minor deformation of the packet as well.
Note that in the illustrated example $\rho =2$ takes a moderate
value and that the approximation in Eq.\ (\ref{appro1}) 
works even better for larger $\rho$ for a limited amount of terms
in the $m$ summation.

An inspection of Fig.\ \ref{Fig3} (a) shows immediately
that for a narrow resonance
the unperturbed incident Gaussian pulse is no good approximation for the
transmitted pulse. The difference is most pronounced for short
times. For example, at  $\tau =1$ a narrow Gaussian 
transmitted peak results in the case of a broad resonance
(Fig.\ \ref{Fig2} (c)) while for the narrow resonance
(Fig.\ \ref{Fig3} (a))  a  much weaker, broader, 
and strongly asymmetric resonance results which 
is nearly completely restricted to negative $q$.
To explain the difference we observe that 
for the broad resonance the
argument of $z$ is smaller than  $3\pi/4$
in the q-region of the transmitted peak
($-5 \leq q \leq 5$ at $\tau =1$,
see Figs.\ \ref{Fig2} (b) and (c)). In contrast,
for the narrow resonance  
the argument  of $z$ is larger than $3\pi/4$ 
in the q-region of the transmitted peak
($q \leq 0$ see  Figs.\ \ref{Fig3} (a) and (c)).
To obtain for small $\rho$ an approximate expression for
$q \leq 0$
one expands $\mbox{erfc} (-z)$ according to \cite{abramowitz}
instead of  $\mbox{erfc} (z)$ as has been
done for the broad resonance.
Then, a superposition of {\em two} factors
\begin{equation}
\psi(x,t) = \psi_{DG} (x,t) + \psi_{DS}(x,t),
\label{rep}
\end{equation}
is obtained with
\begin{equation}
\psi_{DS}(x,t>0)  =  \psi_0 S_0  \rho
\sqrt{2\pi} \exp{ (\beta \rho^2)}
\exp{ \left[ {\Gamma_k \over 2} (x-v_0 t) \right] } 
\exp{\left(i k_0 x - i {\hbar k_0^2 \over 2m} t \right)}.
\label{decay}
\end{equation}
For the broad resonance the factor $\psi_{DS}(x,t>0)$ 
is the dominant contribution to the signal and $\psi_{DG} (x,t)$
is small.
The dominant factor $\psi_{DS}$ results
from a decaying state (DS). This can be seen, first, from
the exponential decay $\propto \exp{ (-\Gamma t /2) }$ 
with $\Gamma = v_0 \Gamma_k$
at a fixed space coordinate $x$. 
Second,
at a fixed time the wave function grows $\propto \exp{ (\Gamma_k x/2) }$
which is typical for a DS \cite{perelomov}.
This growth continues up to $q \approx 0$ where the wave front
of the decaying state signal is located. As can be gathered from
Fig.\ \ref{Fig3} (b) this wave front 
can well be described by replacing in Eq.\ (\ref{centra})
the complementary error function by its small $z$-expansion
$1 - 2z/\sqrt{\pi}+...$.

\begin{figure}[t]
\begin{center}
\noindent \includegraphics*[width=4.0in]{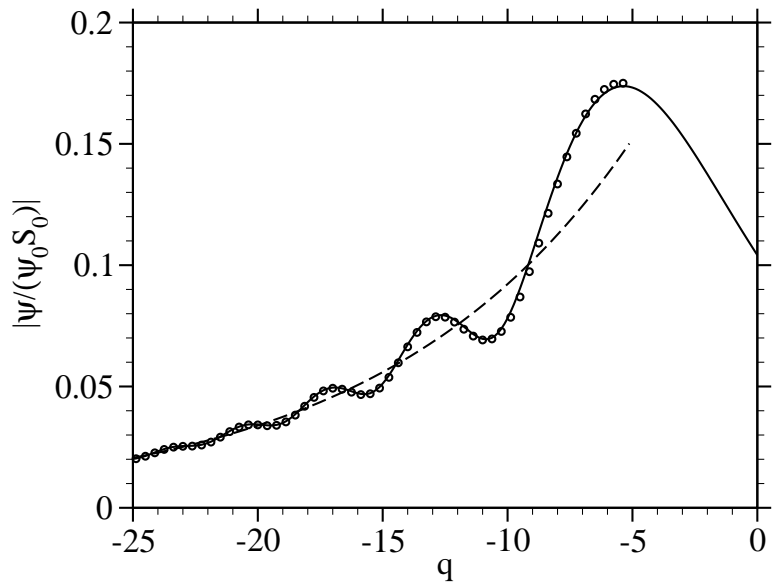}
\end{center}
\caption{Parameters like for the solid line in Fig.\ 3(a) with
$\tau =10$.
In solid line detail of the corresponding curve in Fig.\ 3(a),
dashed line $|\psi_{DS}/(\psi_0 S_0)|$ and in circles the result
of the approximation $|\psi| = |\psi_{DS}| + \psi_{OSZ}$
with $ \psi_{OSZ}$ as given in Eq.\ (\ref{final}).}
 \label{Fig4}
 \end{figure}

In Fig.\ \ref{Fig3} (a) the curve for  $\tau= 10$ 
shows characteristic
oscillations  superimposed to the
decaying state signal. These oscillations are presented in
more detail in Fig.\ \ref{Fig4}. For $q \leq -8$ the exact result
agrees within plot resolution
with the approximation of Eqs.\ (\ref{rep}), Eq.\ (\ref{decay}),
and Eq.\ (\ref{appro1}) (in the latter equation the sum over $m$ 
can be neglected).
The characteristic oscillations stem from the interference between the
dominant  component $\psi_{DS}$ (dashed line in  Fig.\ \ref{Fig4})
and the small component $\psi_{DG}$. 
They are caused by the oscillating
phase in the Gaussian factor  $\exp{[ - q^2/(4  \beta ) ]}$
in $\psi_{DG}$ at complex $\beta$. To show this
we introduce further approximations in Eq. \ (\ref{appro1}): 
first, for moderate times $\tau \sim 10$ one
may set $\beta \sim i \tau /2$
outside the
Gaussian factor  $\exp{[ - q^2/(4  \beta ) ]}$ and, second,
for small $\rho$ the contribution $\rho \sqrt{\beta}$ in $z$
can be omitted.  It results that
\begin{equation}
\psi_{DG} = S_0 \psi_0 {\sqrt{\tau} \over q}  \rho
\exp{ \left(-{q^2 \over 4\beta} + i {\pi \over 4} \right)}
\exp{\left[i k_0 x - i {\hbar k_0^2 \over 2m} t \right]}.
\label{small}
\end{equation}
With this approximation for $\psi_{DG}$ and with Eq.\ (\ref{decay})
a representation
$|\psi(x,t)| = |\psi_{DS}(x,t)| + \psi_{OSZ}(x,t)$ follows
where
\begin{equation}
\psi_{OSZ} = {\mbox{Re} (\psi_{DG} \psi_{DS}^*) \over | \psi_{DS}|} 
 =
 - |\psi_0 S_0| \rho { \sqrt{\tau} \over q} 
\exp{ \left( - {q^2 \over 2 (1 + \tau^2)} \right)}
\sin{ \left( {q^2 \over 2\tau} - {\tau \over 2 } \rho^2 - {\pi \over 4}
      \right)}.
\label{final}
\end{equation}
As shown in in Fig.\ \ref{Fig4} this expression
describes the characteristic oscillations very accurately.
Apart from the constant phase $-\pi /4$
the argument in the sine-factor in Eq.\ (\ref{final})
stems from $\exp{[ - q^2/(4  \beta ) ]}$ in Eq.\ (\ref{small}).
This argument determines solely
the  phase and period of the characteristic oscillations.

\section{Conclusions}

We have analyzed an analytical model for the resonant transmission
of a wave packet. In the limit of a broad
resonance a weakly distorted transmitted pulse results.
In the opposite limit of a narrow resonance the transmitted wave 
is dominated by
a decaying state and characteristic oscillations result.
Our considerations can easily be extended to  
the case of a general Fano resonance.  
The position of the  transmitted peaks at
 different time moments are determined by specific relations  between
 $k_0, k_1 $ and $\Gamma_k$. An understanding of these
relations opens the possibility to solve an
 inverse problem in which the resonance characteristics 
are expressed in terms of the properties of the transmitted pulse.

We acknowledge very helpful discussions with D. Robaschik.

\bibliography{gesamt}
\end{document}